\newif\iffigs\figstrue

\documentstyle[12pt]{article}

\iffigs
  \input epsf
\else
\fi

\batchmode
  \newfont{\footscrfont}{rsfs10}
  \newfont{\footbbbfont}{msbm10}
\errorstopmode

\newif\ifscrf\scrftrue
\ifx\footscrfont\nullfont
  \scrffalse
\fi

\newif\ifamsf\amsftrue
\ifx\footbbbfont\nullfont
  \amsffalse
\fi

\def\ppnumber{\vbox{\baselineskip14pt\hbox{CLNS-95/1359}
\hbox{hep-th/9508154}}}
\def\ppdate{\today}
\def\pplogo{\vbox{\kern-\headheight\kern -15pt
\halign{##&##\hfil\cr&{
\ppnumber}\cr\rule{0pt}{2.5ex}&\ppdate\cr}
}}

\makeatletter
\date{}
\def\dedicatory#1{\def\@date{\normalsize\it#1}}
\def\subjclass#1{\def\@thefnmark{}\@footnotetext{1991
    {\it Mathematics Subject Classification.} #1}}
\def\keywords#1{\def\@thefnmark{}\@footnotetext{
    {\it Key words and phrases.} #1}}

\def\ps@firstpage{\ps@empty \def\@oddhead{\hss\pplogo}%
  \let\@evenhead\@oddhead 
}
\def\maketitle{\par
 \begingroup
 \def\thefootnote{\fnsymbol{footnote}}
 \def\@makefnmark{\hbox
 to 0pt{$^{\@thefnmark}$\hss}}
 \if@twocolumn
 \twocolumn[\@maketitle]
 \else \newpage
 \global\@topnum\z@ \@maketitle \fi\thispagestyle{firstpage}\@thanks
 \endgroup
 \setcounter{footnote}{0}
 \let\maketitle\relax
 \let\@maketitle\relax
 \gdef\@thanks{}\gdef\@author{}\gdef\@title{}\let\thanks\relax}

\def\abstract{\if@twocolumn
\section*{Abstract}
\else \small
\begin{center}
{\bf ABSTRACT}
\end{center}
\quotation
\fi}

\newif\iffn\fnfalse

\@ifundefined{reset@font}{\let\reset@font\empty}{} 
\long\def\@footnotetext#1{\insert\footins{\reset@font\footnotesize
    \interlinepenalty\interfootnotelinepenalty
    \splittopskip\footnotesep
    \splitmaxdepth \dp\strutbox \floatingpenalty \@MM
    \hsize\columnwidth \@parboxrestore
   \edef\@currentlabel{\csname p@footnote\endcsname\@thefnmark}\@makefntext
    {\rule{\z@}{\footnotesep}\ignorespaces
      \fntrue#1\fnfalse\strut}}}
\makeatother

\ifamsf
  \newfont{\bigbbbfont}{msbm10 scaled\magstep2}
  \newfont{\bbbfont}{msbm10 scaled\magstep1}  
  \newfont{\smallbbbfont}{msbm8}
  \newfont{\tinybbbfont}{msbm6}
  \newfont{\smallfootbbbfont}{msbm7}
  \newfont{\tinyfootbbbfont}{msbm5}
\fi

\ifscrf
  \newfont{\scrfont}{rsfs10 scaled\magstep1}  
  \newfont{\smallscrfont}{rsfs7}
  \newfont{\tinyscrfont}{rsfs7}
  \newfont{\smallfootscrfont}{rsfs7}
  \newfont{\tinyfootscrfont}{rsfs7}
\fi

\ifamsf
  \newcommand{\Bbb}[1]{\iffn
      \mathchoice{\mbox{\footbbbfont #1}}{\mbox{\footbbbfont #1}}
      {\mbox{\smallfootbbbfont #1}}{\mbox{\tinyfootbbbfont #1}}\else
      \mathchoice{\mbox{\bbbfont #1}}{\mbox{\bbbfont #1}}
      {\mbox{\smallbbbfont #1}}{\mbox{\tinybbbfont #1}}\fi}
\else
  \def\bigbbbfont{\bf}
  \def\Bbb{\bf}
\fi

\ifscrf
  \newcommand{\Scr}[1]{\iffn
    \mathchoice{\mbox{\footscrfont #1}}{\mbox{\footscrfont #1}}
    {\mbox{\smallfootscrfont #1}}{\mbox{\tinyfootscrfont #1}}\else
    \mathchoice{\mbox{\scrfont #1}}{\mbox{\scrfont #1}}
    {\mbox{\smallscrfont #1}}{\mbox{\tinyscrfont #1}}\fi}
\else
  \def\Scr{\cal}
\fi

\def\R{{\Bbb R}}
\def\Z{{\Bbb Z}}

\def\opeq#1{\advance\lineskip#1 \advance\baselineskip#1
	\advance\lineskiplimit#1}

\def\eqalign#1{\null\,\vcenter{\opeq{2.5\jot}\mathsurround=0pt
	\everycr={}\tabskip=0pt
	\halign{\strut\hfil$\displaystyle{##}$&$\displaystyle{{}##}$\hfil
	\crcr#1\crcr}}\,\null}

\def\cM{{\Scr M}}

\def\cD{{\Scr D}}
\def\cT{{\Scr T}}

\def\cMc{{\hfuzz=100cm\hbox to 0pt{$\;\overline{\phantom{X}}$}\cM}}
\def\barcD{{\hfuzz=100cm\hbox to 0pt{$\;\overline{\phantom{X}}$}\cD}}

\def\ff#1#2{{\textstyle\frac{#1}{#2}}}

\def\RoR{$R\leftrightarrow1/R$}

\ifamsf

\else

\fi

\begin{document}
\setcounter{page}0
\title{\LARGE Some Relationships Between\\ Dualities in String Theory\\[10mm]
\insert\footins{\hbox to\hsize{\footnotesize
Talk given at ``S-Duality and Mirror Symmetry'', Trieste June 1995.\hfil}}}
\author{
Paul S. Aspinwall\\[0.7cm]
\normalsize F.R.~Newman Lab.~of Nuclear Studies,\\
\normalsize Cornell University,\\
\normalsize Ithaca, NY 14853\\[10mm]
}

{\hfuzz=10cm\maketitle}

\def\Large{\large}
\def\LARGE{\large\bf}

\vskip 1.5cm
\vskip 1cm

\begin{abstract}

Some relationships between string theories and eleven-dimensional
supergravity are discussed and reviewed. We see how some relationships
can be derived from others. The cases of $N=2$ supersymmetry in nine
dimensions and $N=4$ supersymmetry in four dimensions are discussed in
some detail. The latter case leads to consideration of quotients of a
K3 surface times a torus and to a possible peculiar relationship
between eleven-dimensional supergravity and the heterotic strings in
ten dimensions.

\end{abstract}

\vfil\break

\section{Introduction}		\label{s:intro}

Recent ideas concerning duality in string theory (such as
\cite{Sen:4d,HT:unity,W:dyn}) have given hope to gaining insights
into some non-perturbative form of string theory. Given the current
status of string theory it is not easy to see how to prove such
statements about duality. Rather, one can take the attitude that such
dualities could be used, in part, as a defining property of string
theory.

Given the many dualities that have been proposed, if we want to
understand how to formulate a new form of string theory, it is
important to know which dualities can be derived from the others.
In particular we appear to have many forms of dualities relating
theories with $N$ supersymmetries in $d$-dimensional flat space-time
for various values of $N$ and $d$. In this talk I will give some simple
ideas on how to formulate relationships
between dualities by concentrating on the cases $N=2,d=9$ and
$N=4,d=4$.

Much of this talk is not original and draws particularly
heavily from \cite{W:dyn} and the later sections are based on the
collaborative work of \cite{AM:Ud}.
Many aspects of section \ref{s:9d} were discussed in \cite{BHO:d=9}
although not in quite the same way as here. The way that the
$U$-duality group is built up in section \ref{s:U} is very similar in
spirit to the work of \cite{Sen:3d}.
It is hoped that the simple
examples explained below show how the dualities can be directly
related to each other in some contexts to build up a rather intricate
picture.

In section \ref{s:dual} we will review the basic dualities used in
this talk. In section \ref{s:9d} we do a ``warm-up'' exercise for the
later sections. In section \ref{s:4d} we have an overview of four
dimensional theories and in section \ref{s:U} we look at the simplest
example of an $N=4$ theory in four dimensions.


\section{Dualities}  	\label{s:dual}

``Duality'' is a much over-used word in the context we wish to use it
and we need to refine our definitions somewhat. Firstly let us discuss
$U$-duality as discussed in \cite{HT:unity}. Consider a particular
string theory. Such a theory will have some deformations (e.g., ``truly
marginal operators'' in the language of conformal field theory) which
will allow us to smoothly reach other string theories. Let us use
$\cM$ to denote the moduli space of such theories. To avoid
complications we will allow $\cM$ to include boundary points a finite
distance away, but we will
not allow ourselves to pass through the boundary to other theories by
processes such as the one described in \cite{GMS:con}. In simple cases
one expects the moduli space to appear naturally in the form
\begin{equation}
  \cM = U\backslash \cT,
\end{equation}
where $\cT\,$ is some smooth domain and $U$ is some discrete group
acting upon it. $\cT\,$ is some generalized notion of a
Teichm\"uller space and $U$ is the group of $U$-dualities. We divide
by discrete groups from the left as $\cT\,$ will typically be a
right-coset as we will see later.

In general one can expect $U$ to be generated by 3 subsets defined
roughly as follows:
\begin{enumerate}
\item $C$-dualities: (This is not conventional notation.) These are
equivalences in $\cT\,$ coming from the classical modular group. That is, if
we can associate our string theory with some geometry, the classical
moduli space of the geometric object will be $C\backslash\cT$. The
canonical example is $Sl(2,\Z)$ for the moduli space of complex
structures on a 2-torus.
\item $T$-dualities: These are further identifications within $\cT\,$
due to the conformal field theories associated with two different
geometries being isomorphic. The canonical example is \RoR\
duality. In some conventions $T$ is a group that contains $C$.
\item $S$-dualities: These are further identifications due to the
effective quantum field theories associated to the string target space
for two apparantly different models being isomorphic. The canonical example
is strong-weak string-coupling duality.
\end{enumerate}

It is generally hoped that the full group $U$ is generated completely
by the elements of $C$, $T$ and $S$.

Together with the notion of $U$-dualities we also have the concept of
equivalences between theories which, at first sight, are qualitatively
different. We list the ones needed in this talk below.

\begin{enumerate}
\item String-string duality. The type IIA superstring compactified on
a K3 surface is equivalent to the heterotic string compactified on a
4-torus. We will denote this relationship by
\begin{equation}
  ({\rm IIA} \to {\rm K3}) \cong ({\rm Het} \to T^4).
\end{equation}
This notion goes back as far as \cite{Sei:K3} but has been developed
subsequently in many other references. The strongly coupled type II
string corresponds to the weakly coupled heterotic string.
\item 11-dimensional supergravity as a string theory
\cite{T:11d,W:dyn}.
\begin{equation}
  ({\rm 11d} \to S^1) \cong {\rm IIA}.	\label{eq:11d}
\end{equation}
In this case the string coupling of the type II string becomes larger
as the radius of the $S^1$ becomes larger.
\item Type II equivalences \cite{DHS:IIAB,DLP:IIAB}.
\begin{equation}
  ({\rm IIA} \to S^1) \cong ({\rm IIB} \to S^1).
\end{equation}
In this case there is an \RoR\ relationship between the two $S^1$s.
\item Heterotic equivalences \cite{N:torus,Gins:torus}.
\begin{equation}
  ({\rm Het}_{E_8\times E_8} \to S^1)\cong ({\rm Het}_{SO(32)}\to
S^1).
\end{equation}
In this case the two 10-dimensional heterotic strings are different
limits in the space $O(1,17)/O(17)$.
\end{enumerate}
It is fairly clear what is meant by each of the above equivalences
with the exception of that for equation (\ref{eq:11d}). Are we really
meant to believe that eleven-dimensional supergravity on a circle
is entirely equivalent to string theory? The answer to this question
is probably no. In \cite{W:dyn} this equivalence was more
conservatively given as that between low-energy effective actions. We
should be aware of this uncertainty whenever eleven-dimensional
supergravity is mentioned below.


\section{Nine Dimensions}  	\label{s:9d}

We may now try to mix the ideas of $U$-duality and equivalences from
the previous section. Consider the case of eleven-dimensional
supergravity compactified on a 2-torus, $T^2$. From the last section
we therefore have
\begin{equation}
  \eqalign{({\rm 11d}\to T^2) &\cong \left[({\rm 11d}\to S^1)
		\to S^1\right]\cr
    &\cong ({\rm IIA}\to S^1)\cr
    &\cong ({\rm IIB}\to S^1),\cr}
\end{equation}
thus relating eleven-dimensional supergravity to the IIB superstring.

Now let $T^2$ be given by a fundamental region in $\R^2$ in the usual
way of the form of a rectangle with sides $r_1$ and $r_2$. The
starting point for our space of theories is thus a quadrant of
$\R^2$. Since the interchange of $r_1$ and $r_2$ clearly has no effect
on the underlying theory, we should divide out by this
interchange. This leads to an infinite triangle as shown in figure
\ref{fig:a}.

\iffigs
\begin{figure}
  \centerline{\epsfxsize=9cm\epsfbox{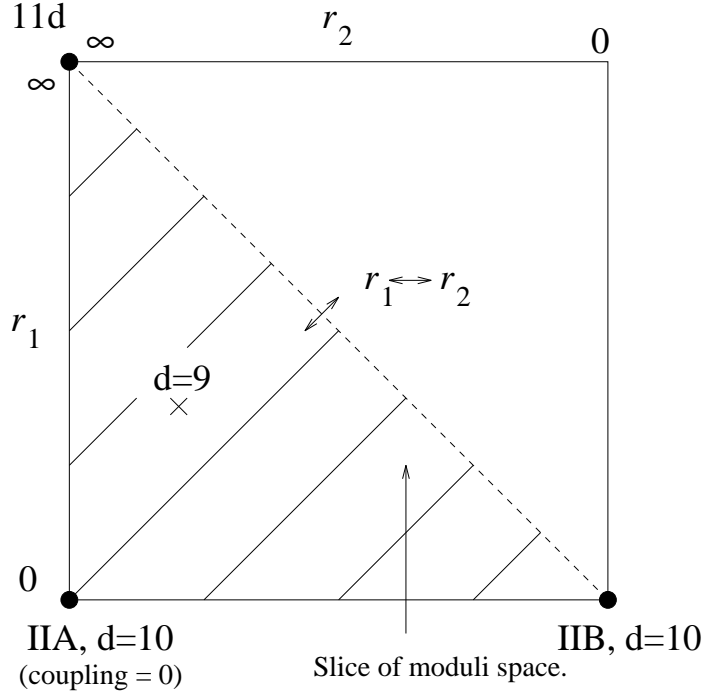}}
  \caption{A slice of the space of theories in 9 dimensions.}
  \label{fig:a}
\end{figure}
\fi

A generic point in this space corresponds to a
nine-dimensional theory. When both radii go to infinity we obtain the
eleven-dimensional theory. Consider the case when $r_1$ is finite and
$r_2$ is infinite. This gives us the correspondence with the IIA
theory as explained in \cite{W:dyn}. Let us denote the IIA string
coupling by $\lambda_A=\exp(\phi_A)$, where $\phi$ is the string
dilaton. We then have
\begin{equation}
  \lambda_A = r_1^{\frac32}.
\end{equation}
Thus the bottom left corner of figure \ref{fig:a} is the type IIA
string in ten dimensions at zero string coupling.

An important point of \cite{W:dyn} is that the ten space-time
dimensions as seen by eleven-dimensional supergravity compactified on
a circle are not quite the same ten space-time dimensions as seen by
the type IIA superstring. They are related by a rescaling. This means
that when $r_2$ is finite, it is not the radius of the circle on which
the type IIA string is compactified. Denoting this latter radius by $r_A$ we
have
\begin{equation}
  r_A = r_1^{\frac12}r_2.
\end{equation}
Now consider the type IIB interpretation. The effective field theories
for the type IIA and IIB theories show that the respective dilatons
must shift when the \RoR\ transformation of \cite{DHS:IIAB,DLP:IIAB}
is performed. This means that we calculate the string coupling of the
type IIB superstring as
\begin{equation}
  \eqalign{
  \lambda_B &= \lambda_A.r_A^{-1}\cr
  &= r_1r_2^{-1}.\cr}
\end{equation}
Consider now the bottom right corner of figure \ref{fig:a}. This now
corresponds to the type IIB string in 10 dimensions but the coupling
is not defined. We should really do a real blow-up at this point do
get the correct moduli space. It is easy to see that the symmetry of
the eleven-dimensional supergravity picture that exchanged the radii
$r_1$ and $r_2$ now translates into the IIB superstring as
\begin{equation}
  \lambda_B \leftrightarrow 1/\lambda_B.
\end{equation}
That is, we have obtained $S$-duality for the IIB string.

Actually, we have not analyzed the complete moduli space. The moduli
space of the torus should also allow the angle between the vectors of
length $r_1$ and $r_2$ to vary. This gives the well-known result that
the moduli space is actually the upper half plane divided by
$Sl(2,\Z)$. In the language of the type IIB superstring, this extra
degree of freedom comes from the expectation value of the axion. Thus,
the $Sl(2,\Z)$ modular invariance of the torus on which
eleven-dimensional supergravity was compactified can be used to ``deduce''
$Sl(2,\Z)$ $S$-duality for the IIB string.

This $S$-duality for the type IIB string was conjectured in
\cite{HT:unity}. We see here that this conjecture is not
independent of the others in the previous section.


\section{Four Dimensional Theories}	\label{s:4d}

Let us consider obtaining four-dimensional theories by compactifying
the known supersymmetric ten-dimensional string theories and
eleven-dimensional
supergravity over manifolds of six and seven dimensions
respectively. The number of supersymmetries in four dimensions can be
found by counting the number of covariantly constant spinors on the six
and seven-dimensional manifolds. This in turn depends purely on the
holonomy group of the compact manifold. In table \ref{t:list} we list
the number of supersymmetries in four dimensions for each higher
dimensional theory.

\begin{table}
\def\st{\rule[-1.5ex]{0em}{4ex}} 
\centerline{\begin{tabular}{|c|c|c|c|}
\hline
\st $N$ & 11d & II & Het \\
\st &Hol,$X$ &Hol,$X$ &Hol,$X$ \\
\hline
\st 8& 1,$T^7$& 1,$T^6$& -- \\
\st 4& $SU(2)$, K3$\times T^3$& $SU(2)$, K3$\times T^2$& 1, $T^6$\\
\st 2& $SU(3)$, CY$\times S^1$& $SU(3)$, CY& $SU(2)$, K3$\times T^2$\\
\st 1& $G_2$, Joyce& --& $SU(3)$, CY\\
\hline
\end{tabular}}
\caption{Four-dimensional theories obtained by compactification.}
\label{t:list}
\end{table}

This table requires some discussion. Firstly we only list possible
geometric compactifications. By using more asymmetric methods, other
models can be built such as an $N=1$ theory built from the type II
string (see, for example, \cite{LNS:asym}). Each of the
entries in the table gives the holonomy of the compact space $X$,
followed by an example of such a space where CY stands for a
Calabi-Yau manifold. A ``Joyce Manifold'' is that of the type
discovered in \cite{Joyce:G2}.

It is tempting to conjecture that for each of the rows in table
\ref{t:list} there is some equivalence between each of the
entries. For the $N=8$ this follows immediately from the conjectured
equivalence between eleven-dimensional supergravity and the type IIA
string by compactifying further on $T^6$. Similarly the $N=4$ row
follows from equivalences mentioned earlier. Analysis on the $N=2$ row
was begun in \cite{KV:N=2,FHSV:N=2}.

In some cases one can classify all the possibilities for $X$ given the
holonomy (see theorem 10.8 of \cite{Sal:hol}).
For the rest of this section we will analyze the case of obtaining
$N=4$ supersymmetry from the type II string where this classification
may be done.  Any 6-dimensional manifold with holonomy $SU(2)$ must be
of the form $({\rm K3}\times T^2)/G$ where $G$ acts freely. Any
element $g\in G$ can be decomposed into an automorphism, $g_1$, of the
K3 surface and an automorphism, $g_2$, of the torus. To retain $SU(2)$
holonomy, these automorphisms must preserve the holomorphic 2-form and
1-form respectively. Such a $g_1$ necessarily has fixed points and so
$g_2$ must act freely. Clearly then, if $g$ is nontrivial, $g_2$ acts
by a translation on the torus.

We can list all possibilities for the group $G$ in this case. Since
any nontrivial element of $G$ must be fixed-point free, the associated
$g_2$ must be nontrivial. Thus $G$ is faithfully represented by
translations in $T^2$. It then follows that $G$ must be of the form
$\Z_m$ or $\Z_m\times\Z_n$ for integers $m,n$. Without loss of
generality we may assume that any element of $G$ acts nontrivially on
the K3 surface. From Nikulin's work \cite{Nik:K3aut} we can then list
the possibilities for $G$. This is done in table \ref{t:K3aut}. $M$ is
the rank of the maximal sublattice of $H^2({\rm K3},\Z)$ that
transforms nontrivially under $G$.

\begin{table}[b]
\centerline{\begin{tabular}{|c||c|c|c|c|c|c|c|c|c|c|c|c|}
\hline
$G$&$\Z_2$&$\Z_2\times\Z_2$&$\Z_2\times\Z_4$&$\Z_2\times\Z_6$
  &$\Z_3$&$\Z_3\times\Z_3$
  &$\Z_4$&$\Z_4\times\Z_4$&$\Z_5$&$\Z_6$&$\Z_7$&$\Z_8$\\
\hline
$M$&8&12&16&18&12&16&14&18&16&16&18&18\\
\hline
\end{tabular}}
\caption{Possible quotienting groups.}
\label{t:K3aut}
\end{table}

Given a type II string compactified on a manifold $X_G=({\rm K3}\times
T^2)/G$, it is natural to ask if this is equivalent to some orbifold
of the heterotic string compactified on $T^6$. That is, can we divide
the $N=4$ row in table \ref{t:list} by $G$ and maintain equivalences?
This appears to be the
case. From \cite{AM:K3p} we expect to identify the lattice of total
cohomology $H^*({\rm K3},\Z)$ with the even-self dual lattice defining
the heterotic string compactified on $T^4$. Thus the action of $G$ on
$H^*({\rm K3},\Z)$ gives us a candidate for an asymmetric orbifold of
the heterotic string. This appears to correspond precisely to the
models studied
in \cite{CHL:bigN,CP:ao}. This point has been investigated further
in \cite{CL:KT/G}.

Two points are worth mentioning. Firstly, the asymmetric orbifolds of
\cite{CP:ao} should provide more examples of heterotic strings than
we have listed here. This is because we have restricted our attention
on the type II side to geometric quotients.
Other models based on type II strings are possible, such as those of
\cite{FK:fc}.
Secondly when one does an
asymmetric orbifold it is important to check that the level-matching
conditions of \cite{Vafa:tor} are satisfied. Given the values of $M$
in table \ref{t:K3aut} we can check whether this is so for all our
examples. The answer is yes and, at least at first sight, this appears
to be remarkable. This should be contrasted to cases where $N=4$
supersymmetry is broken such as \cite{FHSV:N=2}.


\section{$U$-duality}	\label{s:U}

In this section we will analyze the moduli space of the $N=4$ theories
as discussed in the last section. We will focus on the case of
K3$\times T^2$ but it should be easy to extend this analysis to the
quotients. A conjecture for the form of this moduli space was made in
\cite{HT:unity} by making some assumptions about the soliton
spectrum. Here we will be able to rederive this result without making
any direct reference to solitons but rather using the equivalences we
already listed earlier. One can argue that these equivalences rest on
details of the soliton spectrum and so what we are doing in this
section may be completely equivalent to \cite{HT:unity}. Anyway the
analysis below clearly shows
the interrelation between such conjectures. This argument first
appeared in \cite{AM:Ud} and the reader is referred there for
details. The basic idea will be that we will take the moduli space of
theories and try to identify a boundary. This process is not unique
and the different boundaries will correspond to different
interpretations of the theory. The mathematical principles of this
process are in \cite{BS:corner} but the reader is also referred to
\cite{W:dyn,AM:Ud} for a simpler treatment.

By general arguments from supergravity \cite{deRoo:}, the general form
of the Teichm\"uller space for $N=4$ theories in four dimensions is
\begin{equation}
  \cT \cong \frac{O(6,k)}{O(6)\times O(k)} \times \frac{Sl(2)}{U(1)},
\end{equation}
where $k$ is the number of $N=4$ matter supermultiplets. For ease of
notation, when a coset is written $\ff{a}{b}$, the action is assumed
to be from the right. In this case
$k=22$ (or $k=22-M$ for the cases listed in table \ref{t:K3aut}). To
form the moduli space we need to quotient by some group $U$.

{}From the conjecture concerning the rows of table \ref{t:list}, any
point in $\cM = U\backslash\cT$ can be thought of as a
compactification of eleven-dimensional supergravity, the type IIA or
IIB superstring, or the two heterotic strings. Thus each point has
five interpretations (compared to the three interpretations in figure
\ref{fig:a}). Given any one of these five interpretations we should be
able to find part of the moduli space we already understand
nonperturbatively.

Let us begin with the type IIB string. We will assume that we
understand this theory in the weak-coupling limit, i.e., when
$\lambda_B\to0$.  In this case we
should just recover the Teichm\"uller space for conformal field theories with
target space K3$\times T^2$ \cite{Sei:K3} together with directions in
the moduli
space for deformations of the axion and 48 fields from the R-R sector.
Thus we expect to be able to find a boundary of the form
\begin{equation}
\partial_{\lambda_B\to0}\cT\;\cong\;
\frac{O(4,20)}{O(4)\times O(20)}\times\frac{Sl(2)}{U(1)}
  \times\frac{Sl(2)}{U(1)}\times\R^{49},
	\label{eq:BB}
\end{equation}
for the Teichm\"uller space. This is indeed the case following methods
explained in \cite{W:dyn,AM:Ud}. Now since we know that
$O(4,20;\Z)\times Sl(2,\Z)\times Sl(2,\Z)$ acts on the above boundary,
it must also be a subgroup of $U$.

We also know about another limit of this IIB theory. If we rescale the
$T^2$ part so that its area becomes infinite then we are left with a
type IIB string compactified on a K3 surface. The Teichm\"uller space
we are left with should be that for IIB strings on a K3 surface
\cite{W:dyn} together with the complex structure and $B$-field for
$T^2$ and the remaining R-R moduli. This is of the form
\begin{equation}
\partial_{T^2\to\infty}\cT\;\cong\;
\frac{O(5,21)}{O(5)\times O(21)}\times\frac{Sl(2)}{U(1)}
  \times\R^{26},
\end{equation}
which can also be found as a boundary. In this case we show that
$O(5,21;\Z)\times Sl(2,\Z)\subset U$. The relationship between the
boundaries tells us the way these two subgroups fit together within
$U$. Using methods such as those in \cite{AM:K3p} or \cite{Giv:rep}
one can then show that
\begin{equation}
  U \supseteq O(6,22;\Z) \times Sl(2,\Z),
	\label{eq:U}
\end{equation}
and that the equality must be satisfied if $\cM$ is Hausdorff
\cite{AM:K3p}.

No we have found $U$ we can interpret $\cM$ in terms of the other
strings. Firstly we can find the IIA string interpretation. Going to
the weak-coupling string where we really understand what we are doing,
this theory is mirror to the IIB theory. The action of the mirror
map on (\ref{eq:BB}) is within the $O(4,20)$ factor but it exchanges
the two $Sl(2)$ factors. Thus, the only noticeable effect is on the
r\^ole of the $Sl(2,\Z)$ factor in (\ref{eq:U}). For the IIB string
this factor came from the complex structure of $T^2$ whereas now it
acts as a $T$-duality on
the radius and $B$-field as in \cite{DVV:torus}.

Now consider the heterotic string. In this case we know that weakly
coupled string has a Teichm\"uller space of $O(6,22)/(O(6)\times
O(22))\times \R$ thanks to \cite{N:torus}. This is easy to fit into
the required moduli space. In this case the only extra information
coming from the $U$-duality group is the $Sl(2,\Z)$ factor which forms
the $S$-duality group.

Thus we see that for the type IIA, type IIB and heterotic string the
r\^ole of the $Sl(2,\Z)$ group is of a $T$, $C$ and $S$ duality
respectively. This ``triality of dualities'' generalizes the work of
\cite{Duff:S} (and was independently investigated in \cite{DLR:tri}
and recently in \cite{Kal:tri}).

Lastly we need to fit the interpretation of the four-dimensional model
as a compactification of eleven-dimensional supergravity into the
picture. Eleven-dimensional supergravity does not have any weak
coupling limit since the coupling in the action is fixed. However, we
can take the large radius limit. Actually, there are many large radius
limits which appear to be qualitatively different.

In \cite{W:dyn} it was argued that eleven-dimensional supergravity
compactified on a K3 surface was equivalent to the heterotic string
compactified on a 3-torus. Therefore, we should be able to identify
eleven-dimensional supergravity compactified on K3$\times T^3$. Sure
enough, one of the boundaries of the space is
\begin{equation}
\partial_{R_1\to\infty}\cT\;\cong\;
\frac{O(3,19)}{O(3)\times O(19)}\times\frac{Sl(3)}{SO(3)}
  \times\frac{Sl(2)}{U(1)}\times\R^{69},
\end{equation}
where the first factor can be recognized as the Teichm\"uller space of
Ricci-flat metrics on a K3 surface of fixed volume and the second
factor as the Teichm\"uller space of flat metrics of xied volume on
$T^3$. $R_1$ is
some parameter such that the limit $R_1\to\infty$ takes the volume of
both the K3 surface and the 3-torus to infinity. The other degree of
freedom for the two volumes is in the $Sl(2)/U(1)$ factor. It would
thus appear that the $Sl(2,\Z)$ factor in the $U$-duality group has
yet another meaning for the eleven-dimensional supergravity picture.

It is interesting to
note that as soon as we interpret our moduli space in terms of
eleven-dimensional supergravity we recognize factors in the boundary
which correspond {\em classical\/} moduli spaces rather than the
moduli spaces of conformal field theories we were seeing earlier. This
is intimately connected with the fact that a conformal field theory
moduli space has a classical moduli space on its boundary (from the
$\alpha^\prime\to0$ limit).

Another boundary of the moduli space can be written as
\begin{equation}
\partial_{R_2\to\infty}\cT\;\cong\;
\frac{O(2,18)}{O(2)\times O(18)}\times\frac{Sl(4)}{SO(4)}
  \times\ldots
\end{equation}
The second factor is clearly the space of metrics on $T^4$ but what is
the first factor? It can be written as part of the boundary of the
space of Ricci-flat metrics on a K3 surface and so is the space of
some kind of singular K3. One natural interpretation \cite{AM:Ud} is
that the K3 surface has collapsed in itself and is now like a 3
dimensional object. We denote this as a ``squashed K3''. Thus, our
four-dimensional theory has an interpretation as eleven-dimensional
supergravity compactified on a squashed K3 times $T^4$ (both being at
large radius in the above limit).

Continuing this line of argument we can should be able to squash the
K3 surface down to a 2-dimensional and then a 1-dimensional
object. Actually there are two natural 1-dimensional limits, $\Xi_1$
and $\Xi_2$ depending on how we decompose the moduli space with
respect to the lattice structure preserved by $O(4,20;\Z)$. This is
tied to the fact that there are two even self-dual lattices in 16
dimensions. We now see that our four-dimensional theory can be thought
of as eleven-dimensional supergravity compactified on $\Xi_i\times
T^6$. We already knew however that it could also be thought of as the
heterotic string compactified on $T^6$.

The classical moduli space of
the $T^6$ in the eleven-dimensional picture embeds nicely into the
stringy moduli space of the $T^6$ in the heterotic picture and so it
is tempting to ``cancel'' the two $T^6$'s against each other and make
the bold assertion that the heterotic string in ten dimensions is
equivalent to eleven-dimensional supergravity compactified on
$\Xi_i$. Clearly the two choices of $\Xi_i$ should give the $E_8\times
E_8$ string and the $SO(32)$ string.
Whether this statement only makes sense in some delicate limit or
whether we can really directly analyze compactification on $\Xi_i$
remains unclear. In particular, we have not yet calculated what shape
the $\Xi_i$'s are. Clearly neither is a circle since we already know
this should lead to the type IIA string. Thus, if they exist, they
must be some more complicated 1-skeleton object. Clearly some degree
of complexity is required from them since they contain the information
about the gauge group of the heterotic string.

This prediction of some relation between eleven-dimensional
supergravity and heterotic strings arises in this section in a way
very similar to the way $Sl(2,\Z)$ $S$-dulaity for the type IIB string arose in
section \ref{s:9d}. We see that often a full understanding of
the moduli space of a given theory can tell us interesting things
about the relationships between the associated higher-dimensional theories.


\section*{Note added}

Since this talk was presented further constructions, in some ways
similar to the orbifolds presented
here and in \cite{FHSV:N=2}, have been presented in
\cite{SS:pairs,SV:pairs,VW:pairs,HLS:pairs}. In general the $N<4$ case appears
to be more subtle than the version discussed above. Eleven-dimensional
supergravity compactified on manifolds with $G_2$ holonomy has
recently been discussed in \cite{PT:G2,Ach:G2}. The recent paper
\cite{Sch:d=9} has some overlap with section \ref{s:9d}.


\section*{Acknowledgements}

It is a pleasure to thank D.~Morrison for collaboration on some of the
work presented here. I also thank S.~Chaudhuri,
B.~Greene, D.~Joyce, W.~Lerche and R.~Plesser for useful conversations.
The work of the author is supported by a grant from the National
Science Foundation.

\end{document}